\documentstyle[11pt,paspconf]{article}
\def\edcomment#1{\iffalse\marginpar{\raggedright\sl#1\/}\else\relax\fi}
\reversemarginpar

\begin{document}
\title{The Formation and Evolution of LMC Globular Clusters: the database}
\author{S.F. Beaulieu, R. Elson, G. Gilmore, R.A. Johnson, N. Tanvir}
\affil{Institute of Astronomy, University of Cambridge, Madingley Road CB3 0HA,
England}
\author{B. Santiago}
\affil{Instituto de Fisica, Universidade Federal do Rio Grande do Sul,
Av. Bento Goncalves 9500, Porto Alegre RS, CEP: 91501-970, Brazil}

\begin{abstract}
We present details of the database from a large Cycle 7 HST project to study the 
formation and evolution of rich star clusters in the Large Magellanic 
Cloud (see Elson et al., this volume).  Our data set, which includes NICMOS,
WFPC2 and STIS images of the cores and outer regions of 8 clusters, will
enable us to derive deep luminosity functions for the clusters and to 
investigate the universality of the stellar initial mass function.  We
will look for age spreads in the youngest clusters, quantify the population
of binary stars in the cores of the clusters and at the half-mass radii,
and follow the development of mass segregation.
\end{abstract}


\section{Introduction}

The Large Magellanic Cloud (LMC) is unique in containing massive star clusters 
at all stages of evolution. We are presently obtaining very deep HST images 
of 8 optimally selected clusters. This large observing program 
will enable us to probe the cluster's stellar content down to 0.2 M$_\odot$.

In conjunction with state-of-the-art N-body models, being calculated by the
Institute of Astronomy N-body group, these observations will quantify the origin 
and evolution of rich star clusters in the Milky Way, the LMC and beyond.

In Cycle 7, a total of 95 orbits have been granted for this project. We are 
using WFPC2 (F555W,F814W), NICMOS (all three cameras) (F110W,F160W) and 
STIS (CCD/F28X50LP) instruments in primary mode for each cluster, with parallel
observing of surrounding fields as well, giving a total of 285 orbits of
imaging data.

\section{The LMC clusters}

Our clusters are grouped in four pairs, with ages of $10^{7}-10^{10}$ yr (see
Table 1) and are observed at their core and half-mass radius. They 
are among the richest clusters in the LMC and have masses $\approx 10^{4}$ M$_\odot$.
In the youngest clusters, the fraction of primordial binaries, the stellar initial mass
function (IMF), and the degree of primordial mass segregation will investigate the
process of star formation in a protocluster.
Age spreads and a search for pre-main-sequence stars will reveal both the timescale for 
star formation, which has important implications for its trigger, and whether the low
and high mass stars form sequentially or together.
In the intermediate and old clusters, we will trace the development of mass segregation 
and the binary fraction in the core and at the half-mass radius; binaries play a
crucial role in the structural evolution of a cluster, and in particular, affect the
onset of core collapse.
A comparison of the IMFs in our clusters will help answer the far-reaching question
of whether there is such a thing as a universal IMF.

\begin{table}
\caption{The LMC globular cluster sample}
\begin{center}\scriptsize
\begin{tabular}{llllll}
Cluster & log(age/yr) & \multicolumn{1}{c}{[Fe/H]\tablenotemark{1}} &
\multicolumn{1}{c}{{R$_{g}$\deg}\tablenotemark{2}} &
\multicolumn{1}{c}{R$_{c}^{\prime\prime}$\tablenotemark{3}} & Recent CMD \\
\tableline
NGC 1805 & 7.2 & ... & 3.7 & ... & ... \\
NGC 1818 & 7.3 & -0.8 & 3.4 & 7.9 & Will et al.\ (1995) \\
NGC 1831 & 8.7 & -0.3 & 4.6 & 14.2 & Mateo (1998) \\
NGC 1868 & 8.8 & -0.6 & 5.4 & 5.1 & Corsi et al.\ (1994) \\
NGC 2209 & 9.1 & -1.0 & 5.7 & 20.0 & Corsi et al.\ (1994) \\
H14=SL506 & 9.2 & ... & 4.4 & 10.3 & none available \\
NGC 2210 & 10.1 & -2.2 & 5.0 & 7.1 & Brocato et al.\ (1996) \\
H11=SL868& 10.1 & -2.1 & 5.2 & 19.8 & Mighell et al.\ (1996) \\
\end{tabular}
\end{center}

\tablenotetext{1}{The values of metallicity are quite uncertain; for many of the
clusters widely differing values are quoted in the literature}
\tablenotetext{2}{Distance from the LMC centre}
\tablenotetext{3}{Core radius}

\end{table}

In addition to the 8 LMC clusters, we are observing two LMC background fields
in order to correct for contamination from LMC field stars and unresolved
background galaxies. Also from these background fields and the parallel data,
we will study the LMC field star population and the star formation history. 
We are also observing one galactic globular cluster of
similar metallicity to the LMC clusters, NGC 6553, in order to provide an
empirical calibration of the NICMOS and STIS passbands relative to absolute
V magnitude and mass. We have similar data for 47 Tuc from the HST project
7419 (PI, R.F.G. Wyse). 

The data reduction and analysis is performed using STSDAS and photometry 
is performed using the DAOPHOT package.
Some preliminary results can be found in Elson et al. (this volume),
Johnson et al. (1998), and at our webpage: http://www.ast.cam.ac.uk/LMC.

\end{document}